\documentclass[11pt]{article}

    \usepackage[breakable]{tcolorbox}
    \usepackage{parskip} 
    
    \usepackage{iftex}
    \ifPDFTeX
    	\usepackage[T1]{fontenc}
    	\usepackage{mathpazo}
    \else
    	\usepackage{fontspec}
    \fi

    \usepackage{graphicx}
    
    \usepackage{caption}
    \DeclareCaptionFormat{nocaption}{}
    \usepackage[Export]{adjustbox} 
    \adjustboxset{max size={0.9\linewidth}{0.9\paperheight}}
    \usepackage{float}
    \usepackage{xcolor} 
    \usepackage{enumerate} 
    \usepackage{geometry} 
    \usepackage{amsmath} 
    \usepackage{amssymb} 
    \usepackage{textcomp} 
    \AtBeginDocument{%
        \def\PYZsq{\textquotesingle}
    }
    \usepackage{upquote} 
    \usepackage{eurosym} 
    \usepackage[mathletters]{ucs} 
    \usepackage{fancyvrb} 
    \usepackage{grffile} 
    \makeatletter 
    \def\Gread@@xetex#1{%
      \IfFileExists{"\Gin@base".bb}%
      {\Gread@eps{\Gin@base.bb}}%
      {\Gread@@xetex@aux#1}%
    }
    \makeatother

    \usepackage{hyperref}
    \usepackage{titling}
    \usepackage{longtable} 
    \usepackage{booktabs}  
    \usepackage[inline]{enumitem} 
    \usepackage[normalem]{ulem} 
    \usepackage{mathrsfs}

    \definecolor{urlcolor}{rgb}{0,.145,.698}
    \definecolor{linkcolor}{rgb}{.71,0.21,0.01}
    \definecolor{citecolor}{rgb}{.12,.54,.11}

    \definecolor{ansi-black}{HTML}{3E424D}
    \definecolor{ansi-black-intense}{HTML}{282C36}
    \definecolor{ansi-red}{HTML}{E75C58}
    \definecolor{ansi-red-intense}{HTML}{B22B31}
    \definecolor{ansi-green}{HTML}{00A250}
    \definecolor{ansi-green-intense}{HTML}{007427}
    \definecolor{ansi-yellow}{HTML}{DDB62B}
    \definecolor{ansi-yellow-intense}{HTML}{B27D12}
    \definecolor{ansi-blue}{HTML}{208FFB}
    \definecolor{ansi-blue-intense}{HTML}{0065CA}
    \definecolor{ansi-magenta}{HTML}{D160C4}
    \definecolor{ansi-magenta-intense}{HTML}{A03196}
    \definecolor{ansi-cyan}{HTML}{60C6C8}
    \definecolor{ansi-cyan-intense}{HTML}{258F8F}
    \definecolor{ansi-white}{HTML}{C5C1B4}
    \definecolor{ansi-white-intense}{HTML}{A1A6B2}
    \definecolor{ansi-default-inverse-fg}{HTML}{FFFFFF}
    \definecolor{ansi-default-inverse-bg}{HTML}{000000}

    \providecommand{\tightlist}{%
      \setlength{\itemsep}{0pt}\setlength{\parskip}{0pt}}
    \DefineVerbatimEnvironment{Highlighting}{Verbatim}{commandchars=\\\{\}}

    \let\Oldtex\TeX
    \let\Oldlatex\LaTeX
    \renewcommand{\TeX}{\textrm{\Oldtex}}
    \renewcommand{\LaTeX}{\textrm{\Oldlatex}}
    \title{manuscript}

\makeatletter
\def\PY@reset{\let\PY@it=\relax \let\PY@bf=\relax%
    \let\PY@ul=\relax \let\PY@tc=\relax%
    \let\PY@bc=\relax \let\PY@ff=\relax}
\def\PY@tok#1{\csname PY@tok@#1\endcsname}
\def\PY@toks#1+{\ifx\relax#1\empty\else%
    \PY@tok{#1}\expandafter\PY@toks\fi}
\def\PY@do#1{\PY@bc{\PY@tc{\PY@ul{%
    \PY@it{\PY@bf{\PY@ff{#1}}}}}}}
\def\PY#1#2{\PY@reset\PY@toks#1+\relax+\PY@do{#2}}

\expandafter\def\csname PY@tok@w\endcsname{\def\PY@tc##1{\textcolor[rgb]{0.73,0.73,0.73}{##1}}}
\expandafter\def\csname PY@tok@c\endcsname{\let\PY@it=\textit\def\PY@tc##1{\textcolor[rgb]{0.25,0.50,0.50}{##1}}}
\expandafter\def\csname PY@tok@cp\endcsname{\def\PY@tc##1{\textcolor[rgb]{0.74,0.48,0.00}{##1}}}
\expandafter\def\csname PY@tok@k\endcsname{\let\PY@bf=\textbf\def\PY@tc##1{\textcolor[rgb]{0.00,0.50,0.00}{##1}}}
\expandafter\def\csname PY@tok@kp\endcsname{\def\PY@tc##1{\textcolor[rgb]{0.00,0.50,0.00}{##1}}}
\expandafter\def\csname PY@tok@kt\endcsname{\def\PY@tc##1{\textcolor[rgb]{0.69,0.00,0.25}{##1}}}
\expandafter\def\csname PY@tok@o\endcsname{\def\PY@tc##1{\textcolor[rgb]{0.40,0.40,0.40}{##1}}}
\expandafter\def\csname PY@tok@ow\endcsname{\let\PY@bf=\textbf\def\PY@tc##1{\textcolor[rgb]{0.67,0.13,1.00}{##1}}}
\expandafter\def\csname PY@tok@nb\endcsname{\def\PY@tc##1{\textcolor[rgb]{0.00,0.50,0.00}{##1}}}
\expandafter\def\csname PY@tok@nf\endcsname{\def\PY@tc##1{\textcolor[rgb]{0.00,0.00,1.00}{##1}}}
\expandafter\def\csname PY@tok@nc\endcsname{\let\PY@bf=\textbf\def\PY@tc##1{\textcolor[rgb]{0.00,0.00,1.00}{##1}}}
\expandafter\def\csname PY@tok@nn\endcsname{\let\PY@bf=\textbf\def\PY@tc##1{\textcolor[rgb]{0.00,0.00,1.00}{##1}}}
\expandafter\def\csname PY@tok@ne\endcsname{\let\PY@bf=\textbf\def\PY@tc##1{\textcolor[rgb]{0.82,0.25,0.23}{##1}}}
\expandafter\def\csname PY@tok@nv\endcsname{\def\PY@tc##1{\textcolor[rgb]{0.10,0.09,0.49}{##1}}}
\expandafter\def\csname PY@tok@no\endcsname{\def\PY@tc##1{\textcolor[rgb]{0.53,0.00,0.00}{##1}}}
\expandafter\def\csname PY@tok@nl\endcsname{\def\PY@tc##1{\textcolor[rgb]{0.63,0.63,0.00}{##1}}}
\expandafter\def\csname PY@tok@ni\endcsname{\let\PY@bf=\textbf\def\PY@tc##1{\textcolor[rgb]{0.60,0.60,0.60}{##1}}}
\expandafter\def\csname PY@tok@na\endcsname{\def\PY@tc##1{\textcolor[rgb]{0.49,0.56,0.16}{##1}}}
\expandafter\def\csname PY@tok@nt\endcsname{\let\PY@bf=\textbf\def\PY@tc##1{\textcolor[rgb]{0.00,0.50,0.00}{##1}}}
\expandafter\def\csname PY@tok@nd\endcsname{\def\PY@tc##1{\textcolor[rgb]{0.67,0.13,1.00}{##1}}}
\expandafter\def\csname PY@tok@s\endcsname{\def\PY@tc##1{\textcolor[rgb]{0.73,0.13,0.13}{##1}}}
\expandafter\def\csname PY@tok@sd\endcsname{\let\PY@it=\textit\def\PY@tc##1{\textcolor[rgb]{0.73,0.13,0.13}{##1}}}
\expandafter\def\csname PY@tok@si\endcsname{\let\PY@bf=\textbf\def\PY@tc##1{\textcolor[rgb]{0.73,0.40,0.53}{##1}}}
\expandafter\def\csname PY@tok@se\endcsname{\let\PY@bf=\textbf\def\PY@tc##1{\textcolor[rgb]{0.73,0.40,0.13}{##1}}}
\expandafter\def\csname PY@tok@sr\endcsname{\def\PY@tc##1{\textcolor[rgb]{0.73,0.40,0.53}{##1}}}
\expandafter\def\csname PY@tok@ss\endcsname{\def\PY@tc##1{\textcolor[rgb]{0.10,0.09,0.49}{##1}}}
\expandafter\def\csname PY@tok@sx\endcsname{\def\PY@tc##1{\textcolor[rgb]{0.00,0.50,0.00}{##1}}}
\expandafter\def\csname PY@tok@m\endcsname{\def\PY@tc##1{\textcolor[rgb]{0.40,0.40,0.40}{##1}}}
\expandafter\def\csname PY@tok@gh\endcsname{\let\PY@bf=\textbf\def\PY@tc##1{\textcolor[rgb]{0.00,0.00,0.50}{##1}}}
\expandafter\def\csname PY@tok@gu\endcsname{\let\PY@bf=\textbf\def\PY@tc##1{\textcolor[rgb]{0.50,0.00,0.50}{##1}}}
\expandafter\def\csname PY@tok@gd\endcsname{\def\PY@tc##1{\textcolor[rgb]{0.63,0.00,0.00}{##1}}}
\expandafter\def\csname PY@tok@gi\endcsname{\def\PY@tc##1{\textcolor[rgb]{0.00,0.63,0.00}{##1}}}
\expandafter\def\csname PY@tok@gr\endcsname{\def\PY@tc##1{\textcolor[rgb]{1.00,0.00,0.00}{##1}}}
\expandafter\def\csname PY@tok@ge\endcsname{\let\PY@it=\textit}
\expandafter\def\csname PY@tok@gs\endcsname{\let\PY@bf=\textbf}
\expandafter\def\csname PY@tok@gp\endcsname{\let\PY@bf=\textbf\def\PY@tc##1{\textcolor[rgb]{0.00,0.00,0.50}{##1}}}
\expandafter\def\csname PY@tok@go\endcsname{\def\PY@tc##1{\textcolor[rgb]{0.53,0.53,0.53}{##1}}}
\expandafter\def\csname PY@tok@gt\endcsname{\def\PY@tc##1{\textcolor[rgb]{0.00,0.27,0.87}{##1}}}
\expandafter\def\csname PY@tok@err\endcsname{\def\PY@bc##1{\setlength{\fboxsep}{0pt}\fcolorbox[rgb]{1.00,0.00,0.00}{1,1,1}{\strut ##1}}}
\expandafter\def\csname PY@tok@kc\endcsname{\let\PY@bf=\textbf\def\PY@tc##1{\textcolor[rgb]{0.00,0.50,0.00}{##1}}}
\expandafter\def\csname PY@tok@kd\endcsname{\let\PY@bf=\textbf\def\PY@tc##1{\textcolor[rgb]{0.00,0.50,0.00}{##1}}}
\expandafter\def\csname PY@tok@kn\endcsname{\let\PY@bf=\textbf\def\PY@tc##1{\textcolor[rgb]{0.00,0.50,0.00}{##1}}}
\expandafter\def\csname PY@tok@kr\endcsname{\let\PY@bf=\textbf\def\PY@tc##1{\textcolor[rgb]{0.00,0.50,0.00}{##1}}}
\expandafter\def\csname PY@tok@bp\endcsname{\def\PY@tc##1{\textcolor[rgb]{0.00,0.50,0.00}{##1}}}
\expandafter\def\csname PY@tok@fm\endcsname{\def\PY@tc##1{\textcolor[rgb]{0.00,0.00,1.00}{##1}}}
\expandafter\def\csname PY@tok@vc\endcsname{\def\PY@tc##1{\textcolor[rgb]{0.10,0.09,0.49}{##1}}}
\expandafter\def\csname PY@tok@vg\endcsname{\def\PY@tc##1{\textcolor[rgb]{0.10,0.09,0.49}{##1}}}
\expandafter\def\csname PY@tok@vi\endcsname{\def\PY@tc##1{\textcolor[rgb]{0.10,0.09,0.49}{##1}}}
\expandafter\def\csname PY@tok@vm\endcsname{\def\PY@tc##1{\textcolor[rgb]{0.10,0.09,0.49}{##1}}}
\expandafter\def\csname PY@tok@sa\endcsname{\def\PY@tc##1{\textcolor[rgb]{0.73,0.13,0.13}{##1}}}
\expandafter\def\csname PY@tok@sb\endcsname{\def\PY@tc##1{\textcolor[rgb]{0.73,0.13,0.13}{##1}}}
\expandafter\def\csname PY@tok@sc\endcsname{\def\PY@tc##1{\textcolor[rgb]{0.73,0.13,0.13}{##1}}}
\expandafter\def\csname PY@tok@dl\endcsname{\def\PY@tc##1{\textcolor[rgb]{0.73,0.13,0.13}{##1}}}
\expandafter\def\csname PY@tok@s2\endcsname{\def\PY@tc##1{\textcolor[rgb]{0.73,0.13,0.13}{##1}}}
\expandafter\def\csname PY@tok@sh\endcsname{\def\PY@tc##1{\textcolor[rgb]{0.73,0.13,0.13}{##1}}}
\expandafter\def\csname PY@tok@s1\endcsname{\def\PY@tc##1{\textcolor[rgb]{0.73,0.13,0.13}{##1}}}
\expandafter\def\csname PY@tok@mb\endcsname{\def\PY@tc##1{\textcolor[rgb]{0.40,0.40,0.40}{##1}}}
\expandafter\def\csname PY@tok@mf\endcsname{\def\PY@tc##1{\textcolor[rgb]{0.40,0.40,0.40}{##1}}}
\expandafter\def\csname PY@tok@mh\endcsname{\def\PY@tc##1{\textcolor[rgb]{0.40,0.40,0.40}{##1}}}
\expandafter\def\csname PY@tok@mi\endcsname{\def\PY@tc##1{\textcolor[rgb]{0.40,0.40,0.40}{##1}}}
\expandafter\def\csname PY@tok@il\endcsname{\def\PY@tc##1{\textcolor[rgb]{0.40,0.40,0.40}{##1}}}
\expandafter\def\csname PY@tok@mo\endcsname{\def\PY@tc##1{\textcolor[rgb]{0.40,0.40,0.40}{##1}}}
\expandafter\def\csname PY@tok@ch\endcsname{\let\PY@it=\textit\def\PY@tc##1{\textcolor[rgb]{0.25,0.50,0.50}{##1}}}
\expandafter\def\csname PY@tok@cm\endcsname{\let\PY@it=\textit\def\PY@tc##1{\textcolor[rgb]{0.25,0.50,0.50}{##1}}}
\expandafter\def\csname PY@tok@cpf\endcsname{\let\PY@it=\textit\def\PY@tc##1{\textcolor[rgb]{0.25,0.50,0.50}{##1}}}
\expandafter\def\csname PY@tok@c1\endcsname{\let\PY@it=\textit\def\PY@tc##1{\textcolor[rgb]{0.25,0.50,0.50}{##1}}}
\expandafter\def\csname PY@tok@cs\endcsname{\let\PY@it=\textit\def\PY@tc##1{\textcolor[rgb]{0.25,0.50,0.50}{##1}}}

\def\PYZus{\char`\_}
\def\PYZob{\char`\{}
\def\PYZcb{\char`\}}

\def\PYZsh{\char`\#}

\def\PYZhy{\char`\-}
\def\PYZsq{\char`\'}
\def\PYZdq{\char`\"}
\def\PYZti{\char`\~}

\makeatother

    \makeatletter
        \newbox\Wrappedcontinuationbox 
        \newbox\Wrappedvisiblespacebox 
        \newcommand*\Wrappedvisiblespace {\textcolor{red}{\textvisiblespace}} 
        \newcommand*\Wrappedcontinuationsymbol {\textcolor{red}{\llap{\tiny$\m@th\hookrightarrow$}}} 
        \newcommand*\Wrappedcontinuationindent {3ex } 
        \newcommand*\Wrappedafterbreak {\kern\Wrappedcontinuationindent\copy\Wrappedcontinuationbox} 
        \newcommand*\Wrappedbreaksatspecials {%
            \def\PYGZus{\discretionary{\char`\_}{\Wrappedafterbreak}{\char`\_}}%
            \def\PYGZob{\discretionary{}{\Wrappedafterbreak\char`\{}{\char`\{}}%
            \def\PYGZcb{\discretionary{\char`\}}{\Wrappedafterbreak}{\char`\}}}%
            \def\PYGZca{\discretionary{\char`\^}{\Wrappedafterbreak}{\char`\^}}%
            \def\PYGZam{\discretionary{\char`\&}{\Wrappedafterbreak}{\char`\&}}%
            \def\PYGZlt{\discretionary{}{\Wrappedafterbreak\char`\<}{\char`\<}}%
            \def\PYGZgt{\discretionary{\char`\>}{\Wrappedafterbreak}{\char`\>}}%
            \def\PYGZsh{\discretionary{}{\Wrappedafterbreak\char`\#}{\char`\#}}%
            \def\PYGZpc{\discretionary{}{\Wrappedafterbreak\char`\%}{\char`\%}}%
            \def\PYGZdl{\discretionary{}{\Wrappedafterbreak\char`\$}{\char`\$}}%
            \def\PYGZhy{\discretionary{\char`\-}{\Wrappedafterbreak}{\char`\-}}%
            \def\PYGZsq{\discretionary{}{\Wrappedafterbreak\textquotesingle}{\textquotesingle}}%
            \def\PYGZdq{\discretionary{}{\Wrappedafterbreak\char`\"}{\char`\"}}%
            \def\PYGZti{\discretionary{\char`\~}{\Wrappedafterbreak}{\char`\~}}%
        } 
        \newcommand*\Wrappedbreaksatpunct {%
            \lccode`\~`\.\lowercase{\def~}{\discretionary{\hbox{\char`\.}}{\Wrappedafterbreak}{\hbox{\char`\.}}}%
            \lccode`\~`\,\lowercase{\def~}{\discretionary{\hbox{\char`\,}}{\Wrappedafterbreak}{\hbox{\char`\,}}}%
            \lccode`\~`\;\lowercase{\def~}{\discretionary{\hbox{\char`\;}}{\Wrappedafterbreak}{\hbox{\char`\;}}}%
            \lccode`\~`\:\lowercase{\def~}{\discretionary{\hbox{\char`\:}}{\Wrappedafterbreak}{\hbox{\char`\:}}}%
            \lccode`\~`\?\lowercase{\def~}{\discretionary{\hbox{\char`\?}}{\Wrappedafterbreak}{\hbox{\char`\?}}}%
            \lccode`\~`\!\lowercase{\def~}{\discretionary{\hbox{\char`\!}}{\Wrappedafterbreak}{\hbox{\char`\!}}}%
            \lccode`\~`\/\lowercase{\def~}{\discretionary{\hbox{\char`\/}}{\Wrappedafterbreak}{\hbox{\char`\/}}}%
            \catcode`\.\active
            \catcode`\,\active 
            \catcode`\;\active
            \catcode`\:\active
            \catcode`\?\active
            \catcode`\!\active
            \catcode`\/\active 
            \lccode`\~`\~ 	
        }
    \makeatother

    \let\OriginalVerbatim=\Verbatim
    \makeatletter
    \renewcommand{\Verbatim}[1][1]{%
        \sbox\Wrappedcontinuationbox {\Wrappedcontinuationsymbol}%
        \sbox\Wrappedvisiblespacebox {\FV@SetupFont\Wrappedvisiblespace}%
        \def\FancyVerbFormatLine ##1{\hsize\linewidth
            \vtop{\raggedright\hyphenpenalty\z@\exhyphenpenalty\z@
                \doublehyphendemerits\z@\finalhyphendemerits\z@
                \strut ##1\strut}%
        }%
        \def\FV@Space {%
            \nobreak\hskip\z@ plus\fontdimen3\font minus\fontdimen4\font
            \discretionary{\copy\Wrappedvisiblespacebox}{\Wrappedafterbreak}
            {\kern\fontdimen2\font}%
        }%
        
        \Wrappedbreaksatspecials
        \OriginalVerbatim[#1,codes*=\Wrappedbreaksatpunct]%
    }
    \makeatother

    \definecolor{incolor}{HTML}{303F9F}
    \definecolor{outcolor}{HTML}{D84315}
    \definecolor{cellborder}{HTML}{CFCFCF}
    \definecolor{cellbackground}{HTML}{F7F7F7}
    
    \makeatletter
    \newcommand{\boxspacing}{\kern\kvtcb@left@rule\kern\kvtcb@boxsep}
    \makeatother
    \newcommand{\prompt}[4]{
        \ttfamily\llap{{\color{#2}[#3]:\hspace{3pt}#4}}\vspace{-\baselineskip}
    }

    \hypersetup{
      breaklinks=true,  
      colorlinks=true,
      urlcolor=urlcolor,
      linkcolor=linkcolor,
      citecolor=citecolor,
      }
    
\begin{document}
    
    \begin{center}
    	
    ~
    
    \bigskip\bigskip\bigskip\bigskip\bigskip\bigskip
    
    \Large\textbf{Urban Street Network Analysis in a Computational Notebook}\normalsize
    
    \bigskip\bigskip
	
	Geoff Boeing
	
	Department of Urban Planning and Spatial Analysis\\
	Sol Price School of Public Policy\\
	University of Southern California\\
	boeing@usc.edu
	
	\bigskip\bigskip	
	\end{center}
	
	\begin{changemargin}{2cm}{2cm} 
	\small
	\textbf{Abstract:} Computational notebooks offer researchers, practitioners, students, and
	educators the ability to interactively conduct analytics and disseminate
	reproducible workflows that weave together code, visuals, and
	narratives. This article explores the potential of computational
	notebooks in urban analytics and planning, demonstrating their utility
	through a case study of OSMnx and its tutorials repository. OSMnx is a
	Python package for working with OpenStreetMap data and modeling,
	analyzing, and visualizing street networks anywhere in the world. Its
	official demos and tutorials are distributed as open-source Jupyter
	notebooks on GitHub. This article showcases this resource by documenting
	the repository and demonstrating OSMnx interactively through a synoptic
	tutorial adapted from the repository. It illustrates how to download
	urban data and model street networks for various study sites, compute
	network indicators, visualize street centrality, calculate routes, and
	work with other spatial data such as building footprints and points of
	interest. Computational notebooks help introduce methods to new users
	and help researchers reach broader audiences interested in learning
	from, adapting, and remixing their work. Due to their utility and
	versatility, the ongoing adoption of computational notebooks in urban
	planning, analytics, and related geocomputation disciplines should
	continue into the future.\footnote{\textbf{This is a preprint of:} Boeing, G. 2020. Urban Street Network Analysis in a Computational Notebook. \textit{Region} 6 (3), 39--51. doi:10.18335/region.v6i3.278}
	\end{changemargin}

    \hypertarget{introduction}{%
\section{Introduction}\label{introduction}}

A traditional academic and professional divide has long existed between
code creators and code users. The former would develop software tools
and workflows for professional or research applications, which the
latter would then use to conduct analyses or answer scientific
questions. Today, however, these boundary lines increasingly blur as
computation percolates throughout both the natural and social sciences.
As quantitatively-oriented academics gradually shift away from
monolithic, closed-source data analysis software systems like SPSS and
ArcGIS, they increasingly embrace coding languages like R and Python to
script and document their research workflows (Padgham et al.~2019).
Developing shareable, reproducible, and recomputable scripts in R or
Python to acquire, transform, describe, visualize, and model data, these
researchers act as both code creators and code users.

An important trend in this methodological trajectory has been the
widespread adoption of the computational notebook. A computational
notebook is a computer file that replaces the traditional lab notebook
and intersperses plain-language narrative, hyperlinks, and images with
snippets of code in the paradigm of literate programming (Knuth 1992).
These notebooks are easily distributed and integrate well with version
control systems like Git because they are simply structured text files.
They have pedagogical value in introducing students to computational
thinking and coding techniques while thoroughly explaining each new
programming language facet as it is introduced. They also offer research
value in documenting data, questions, hypotheses, procedures,
experiments, and results in detail alongside each's attendant
computations (Pérez and Granger 2007; Kluyver et al.~2016).

Computational notebooks thus open up the world of analytics to a wider
audience than was possible in the past. This particularly impacts
disciplines that encompass diverse methodologies and skillsets. For
example, urban planning, like many academic domains related or adjacent
to regional science, comprises a broad set of scholars, students, and
working professionals with a wide range of computational aptitude. Some
urban planners focus on policymaking within the political constraints of
city hall. Others employ qualitative methods to work in and with
vulnerable communities. Others develop simulation models to forecast
urbanization patterns and infrastructure needs. Others intermingle
these, and many more, different approaches to understanding and shaping
the city. Yet all urban planners benefit from basic quantitative
literacy and an ability to reason critically with data. This scholarly
and professional imperative aligns with the growing importance of
computational thinking in the urban context and parallel trends in
geocomputation (Harris et al.~2017), geographic data science (Kang et
al.~2019; Poorthuis and Zook 2019; Singleton and Arribas-Bel 2019), and
the open-source/open-science movements (Rey 2019).

Urban planning and its related disciplines benefit accordingly from the
growing adoption of computational notebooks in pedagogy, research, and
practice. Computation is increasingly central to the field and its
practitioners benefit from open and reproducible approaches to analyzing
urban data and predicting city futures (Kedron et al.~2019; Kontokosta
2018; Batty 2019). In the Python universe, for example, numerous new
tools now exist to support urban analytics and planning processes,
including data wrangling/analysis (pandas), visualization (matplotlib),
geospatial wrangling/analysis (geopandas), spatial data science and
econometrics (pySAL), mapping (cartopy), web mapping (folium), network
analysis (NetworkX), land use modeling/simulation (UrbanSim),
activity-based travel modeling (ActivitySim), and computational
notebooks themselves (Jupyter).

Another Python tool useful for urban planning research and practice --
and the primary focus of this article -- is OSMnx, a package for street
network analysis (Boeing 2017). OSMnx allows users to download spatial
data (including street networks, other networked infrastructure,
building footprints, and points of interest) from OpenStreetMap then
model, analyze, and visualize them. To introduce new users to its
functionality and capabilities, OSMnx's official demos and tutorials are
developed and maintained in Jupyter notebook format. This repository in
turn offers a compelling case study of the potential of computational
notebooks to document and disseminate geospatial software tools.

This article introduces OSMnx as a computational tool for urban street
network analysis by way of these computational notebooks. It describes
their repository and highlights examples from them, inline here, to
illustrate the use and value of computational notebooks. To do so, it
demonstrates how to interactively execute the code in this article
itself by using Docker to run a containerized computational environment
including Jupyter Lab as an interactive web-based interface. The article
is organized as follows. First, it presents the repository containing
OSMnx's demo and tutorial notebooks. Then it describes how to run
OSMnx's computational environment via Docker. Next it demonstrates the
use of OSMnx interactively in the article itself through a synoptic
tutorial adapted from this repository. Finally, it concludes by
discussing the prospects of notebooks for facilitating the adoption of
computational workflows in urban analytics and planning.

    \hypertarget{the-osmnx-examples-repository}{%
\section{The OSMnx Examples
Repository}\label{the-osmnx-examples-repository}}

OSMnx's official demos, tutorials, and examples are in Jupyter notebook
format in a \href{https://github.com/gboeing/osmnx-examples}{GitHub
repository}. The repository's root contains a license file, a readme
file, an environment definition file, repository contributing
guidelines, and a notebooks folder. Within that folder, the repository
contains 19 thematically organized Jupyter notebook files that
collectively provide a short self-directed tutorial-style course in
using OSMnx. The following notebooks are included there:

\begin{enumerate}
\def\labelenumi{\arabic{enumi}.}
\tightlist
\item
  An introductory survey of features
\item
  A more comprehensive overview of OSMnx's basic functionality
\item
  Using OSMnx to produce shapefiles
\item
  Modeling and visualizing street networks in different places at
  different scales
\item
  Using OSMnx's network topology cleaning and simplification features
\item
  Saving and loading data to/from disk with OSMnx
\item
  Conducting street network analyses with OSMnx and its NetworkX
  dependency
\item
  Visualizing street networks and study sites
\item
  Working with dual graphs of street networks
\item
  Producing figure-ground diagrams for urban form analysis
\item
  Working with building footprints
\item
  Interactive web mapping of street networks and routes
\item
  Attaching elevations to the network and calculating street grades
\item
  Working with isolines and isochrones
\item
  Cleaning complex street intersections
\item
  Calculating street bearings
\item
  Working with other types of spatial infrastructure
\item
  Visualizing street network orientation with polar histograms
\item
  Interfacing between OSMnx and igraph for fast algorithm
  implementations in the C language
\end{enumerate}

This resource is useful for introducing users to the OSMnx software
package, demonstrating how to download, model, analyze, and visualize
street networks in Python, and illustrating several basic and
intermediate spatial network analyses. To run the code examples in this
resource repository, one must have access to a Python installation with
the code dependencies installed, including Jupyter itself for running
the notebook files. Two primary options exist for installing this
computational environment. The first is installing Python locally, then
configuring it and installing all the necessary packages and
dependencies. This can be time-consuming and requires some prior
experience beyond the scope of this article. The second, and easier,
option is to simply run everything in a pre-built Docker container. This
latter option is detailed in the following section.

    \hypertarget{the-computational-environment}{%
\section{The Computational
Environment}\label{the-computational-environment}}

The OSMnx project's reference Docker image contains a stable, consistent
computational environment for running OSMnx on any computer. Docker is a
virtualization tool that allows complex software stacks to be delivered
as self-contained packages called images, allowing users to run software
without having to compile or install a complex chain of dependencies.
Instead, users install Docker on their computer then tell it to run a
certain image as an instance called a container.

This article can be read in its static form (i.e., HTML or PDF) or it
can be executed interactively (i.e., via its .ipynb Jupyter notebook
file). For interactive execution, install Docker and run the official
OSMnx container as follows. First, download and install
\href{https://www.docker.com/products/docker-desktop}{Docker Desktop}.
Once it is installed and running on your computer, open Docker's
settings/preferences and ensure that your local drives are shared with
Docker so the container has access to the notebook file. Then run the
\href{https://hub.docker.com/r/gboeing/osmnx}{OSMnx Docker container}
(which contains a Python installation and all the packages needed to run
OSMnx, including Jupyter Lab) by following the platform-specific
instructions below.

If you are on \emph{Windows} open a command prompt, change directory to
the location of this notebook file then run:

\begin{verbatim}
docker run --rm -it -p 8888:8888 -v "%cd%":/home/jovyan/work gboeing/osmnx:v0.10
\end{verbatim}

If you are on \emph{Mac/Linux} open a terminal window, change directory
to the location of this notebook file then run:

\begin{verbatim}
docker run --rm -it -p 8888:8888 -v "$PWD":/home/jovyan/work gboeing/osmnx:v0.10
\end{verbatim}

Once the container is running per these instructions, open your
computer's web browser and visit \url{http://localhost:8888} to access
Jupyter Lab and open this article's notebook file.

    \hypertarget{street-network-analysis-with-osmnx}{%
\section{Street Network Analysis with
OSMnx}\label{street-network-analysis-with-osmnx}}

Here we showcase the resource repository inline to demonstrate potential
applications. In particular, we highlight specific material from its
notebooks (enumerated above), adapting their code into this interactive
article to introduce OSMnx and illustrate some of the capabilities of a
computational notebook.

First we import the necessary Python modules:

    \begin{tcolorbox}[breakable, size=fbox, boxrule=1pt, pad at break*=1mm,colback=cellbackground, colframe=cellborder]
\prompt{In}{incolor}{1}{\boxspacing}
\begin{Verbatim}[commandchars=\\\{\}]
\PY{k+kn}{import} \PY{n+nn}{matplotlib}\PY{n+nn}{.}\PY{n+nn}{cm} \PY{k}{as} \PY{n+nn}{cm}
\PY{k+kn}{import} \PY{n+nn}{matplotlib}\PY{n+nn}{.}\PY{n+nn}{colors} \PY{k}{as} \PY{n+nn}{colors}
\PY{k+kn}{import} \PY{n+nn}{networkx} \PY{k}{as} \PY{n+nn}{nx}
\PY{k+kn}{from} \PY{n+nn}{IPython}\PY{n+nn}{.}\PY{n+nn}{display} \PY{k}{import} \PY{n}{Image}
\PY{k+kn}{from} \PY{n+nn}{pprint} \PY{k}{import} \PY{n}{pprint}
\end{Verbatim}
\end{tcolorbox}

    matplotlib is a package for data visualization and plotting. NetworkX is
a package for generic network analysis. IPython provides interactive
computing and underpins our Python-language Jupyter environment (Pérez
and Granger 2007). pprint allows us to ``pretty print'' Python data
structures to make them easier to read inline.

Next we import OSMnx itself, configure it, and display its version
number:

    \begin{tcolorbox}[breakable, size=fbox, boxrule=1pt, pad at break*=1mm,colback=cellbackground, colframe=cellborder]
\prompt{In}{incolor}{2}{\boxspacing}
\begin{Verbatim}[commandchars=\\\{\}]
\PY{k+kn}{import} \PY{n+nn}{osmnx} \PY{k}{as} \PY{n+nn}{ox}
\PY{n}{ox}\PY{o}{.}\PY{n}{config}\PY{p}{(}\PY{n}{log\PYZus{}console}\PY{o}{=}\PY{k+kc}{True}\PY{p}{,} \PY{n}{use\PYZus{}cache}\PY{o}{=}\PY{k+kc}{True}\PY{p}{)}
\PY{n}{ox}\PY{o}{.}\PY{n}{\PYZus{}\PYZus{}version\PYZus{}\PYZus{}}
\end{Verbatim}
\end{tcolorbox}

            \begin{tcolorbox}[breakable, size=fbox, boxrule=.5pt, pad at break*=1mm, opacityfill=0]
\prompt{Out}{outcolor}{2}{\boxspacing}
\begin{Verbatim}[commandchars=\\\{\}]
'0.10'
\end{Verbatim}
\end{tcolorbox}
        
    The configuration step tells OSMnx to log its actions to the terminal
window and to use a cache. This cache saves a local copy of any data
downloaded by OSMnx to prevent re-downloading the same data each time
the code is run.

Next we use OSMnx to download the street network of Piedmont,
California, construct a graph model of it (via NetworkX), then plot the
network with the \texttt{plot\_graph} function (which uses matplotlib
under the hood):

    \begin{tcolorbox}[breakable, size=fbox, boxrule=1pt, pad at break*=1mm,colback=cellbackground, colframe=cellborder]
\prompt{In}{incolor}{3}{\boxspacing}
\begin{Verbatim}[commandchars=\\\{\}]
\PY{c+c1}{\PYZsh{} create a graph of Piedmont\PYZsq{}s drivable street network then plot it}
\PY{n}{G} \PY{o}{=} \PY{n}{ox}\PY{o}{.}\PY{n}{graph\PYZus{}from\PYZus{}place}\PY{p}{(}\PY{l+s+s1}{\PYZsq{}}\PY{l+s+s1}{Piedmont, California, USA}\PY{l+s+s1}{\PYZsq{}}\PY{p}{,} \PY{n}{network\PYZus{}type}\PY{o}{=}\PY{l+s+s1}{\PYZsq{}}\PY{l+s+s1}{drive}\PY{l+s+s1}{\PYZsq{}}\PY{p}{)}
\PY{n}{fig}\PY{p}{,} \PY{n}{ax} \PY{o}{=} \PY{n}{ox}\PY{o}{.}\PY{n}{plot\PYZus{}graph}\PY{p}{(}\PY{n}{G}\PY{p}{)}
\end{Verbatim}
\end{tcolorbox}

    \begin{center}
    \adjustimage{max size={0.5\linewidth}{0.5\paperheight}}{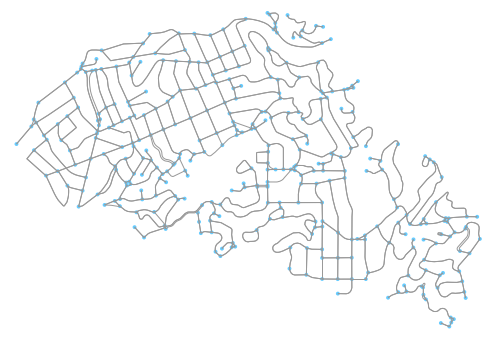}
    \end{center}
    { \hspace*{\fill} \\}
    
    In the figure above, the network's intersections and dead-ends (i.e.,
graph nodes) appear as light blue circles and its street segments (i.e.,
graph edges) appear as gray lines. This is the street network within the
municipal boundaries of the city of Piedmont, California. We select this
study site for pedagogical purposes as it is a relatively small,
self-contained municipality and lends itself to convenient visualization
and indicator calculation here. Note that we specified
\texttt{network\_type=\textquotesingle{}drive\textquotesingle{}} so this
is specifically the drivable network in the city. OSMnx can also
automatically download and model walkable and bikeable street networks
by changing this argument.

\hypertarget{calculating-network-indicators}{%
\subsection{Calculating Network
Indicators}\label{calculating-network-indicators}}

Now that we have a model of the network, we can calculate some
statistics and indicators. First, what area does our network cover in
square meters? To calculate this, we project the graph, convert its
projected nodes to a geopandas GeoDataFrame, then calculate the area of
the convex hull of this set of node points in the Euclidean plane:

    \begin{tcolorbox}[breakable, size=fbox, boxrule=1pt, pad at break*=1mm,colback=cellbackground, colframe=cellborder]
\prompt{In}{incolor}{4}{\boxspacing}
\begin{Verbatim}[commandchars=\\\{\}]
\PY{c+c1}{\PYZsh{} project graph then calculate its nodes\PYZsq{} convex hull area}
\PY{n}{G\PYZus{}proj} \PY{o}{=} \PY{n}{ox}\PY{o}{.}\PY{n}{project\PYZus{}graph}\PY{p}{(}\PY{n}{G}\PY{p}{)}
\PY{n}{nodes\PYZus{}proj} \PY{o}{=} \PY{n}{ox}\PY{o}{.}\PY{n}{graph\PYZus{}to\PYZus{}gdfs}\PY{p}{(}\PY{n}{G\PYZus{}proj}\PY{p}{,} \PY{n}{edges}\PY{o}{=}\PY{k+kc}{False}\PY{p}{)}
\PY{n}{graph\PYZus{}area\PYZus{}m} \PY{o}{=} \PY{n}{nodes\PYZus{}proj}\PY{o}{.}\PY{n}{unary\PYZus{}union}\PY{o}{.}\PY{n}{convex\PYZus{}hull}\PY{o}{.}\PY{n}{area}
\PY{n}{graph\PYZus{}area\PYZus{}m}
\end{Verbatim}
\end{tcolorbox}

            \begin{tcolorbox}[breakable, size=fbox, boxrule=.5pt, pad at break*=1mm, opacityfill=0]
\prompt{Out}{outcolor}{4}{\boxspacing}
\begin{Verbatim}[commandchars=\\\{\}]
4224782.349449131
\end{Verbatim}
\end{tcolorbox}
        
    Thus, this network covers approximately 4.2 square kilometers. When
projecting graphs, OSMnx by default uses the Universal Transverse
Mercator (UTM) coordinate system and automatically determines the UTM
zone for projection based on the network's centroid. Other coordinate
reference systems can be defined by the user to customize this
projection behavior.

Next, we compute and inspect some basic stats about the network:

    \begin{tcolorbox}[breakable, size=fbox, boxrule=1pt, pad at break*=1mm,colback=cellbackground, colframe=cellborder]
\prompt{In}{incolor}{5}{\boxspacing}
\begin{Verbatim}[commandchars=\\\{\}]
\PY{c+c1}{\PYZsh{} calculate and print basic network stats}
\PY{n}{stats} \PY{o}{=} \PY{n}{ox}\PY{o}{.}\PY{n}{basic\PYZus{}stats}\PY{p}{(}\PY{n}{G\PYZus{}proj}\PY{p}{,} \PY{n}{area}\PY{o}{=}\PY{n}{graph\PYZus{}area\PYZus{}m}\PY{p}{,} \PY{n}{clean\PYZus{}intersects}\PY{o}{=}\PY{k+kc}{True}\PY{p}{,} \PY{n}{circuity\PYZus{}dist}\PY{o}{=}\PY{l+s+s1}{\PYZsq{}}\PY{l+s+s1}{euclidean}\PY{l+s+s1}{\PYZsq{}}\PY{p}{)}
\PY{n}{pprint}\PY{p}{(}\PY{n}{stats}\PY{p}{)}
\end{Verbatim}
\end{tcolorbox}

    \begin{Verbatim}[commandchars=\\\{\}]
\{'circuity\_avg': 1.11354525174028,
 'clean\_intersection\_count': 271,
 'clean\_intersection\_density\_km': 64.1453162753664,
 'edge\_density\_km': 26951.828421373437,
 'edge\_length\_avg': 121.39190724946685,
 'edge\_length\_total': 113865.60899999991,
 'intersection\_count': 312,
 'intersection\_density\_km': 73.84995822108604,
 'k\_avg': 5.421965317919075,
 'm': 938,
 'n': 346,
 'node\_density\_km': 81.89771007851208,
 'self\_loop\_proportion': 0.006396588486140725,
 'street\_density\_km': 14061.652905680734,
 'street\_length\_avg': 121.23963877551029,
 'street\_length\_total': 59407.42300000004,
 'street\_segments\_count': 490,
 'streets\_per\_node\_avg': 2.953757225433526,
 'streets\_per\_node\_counts': \{0: 0, 1: 34, 2: 0, 3: 263, 4: 47, 5: 1, 6: 1\},
 'streets\_per\_node\_proportion': \{0: 0.0,
                                 1: 0.09826589595375723,
                                 2: 0.0,
                                 3: 0.7601156069364162,
                                 4: 0.13583815028901733,
                                 5: 0.002890173410404624,
                                 6: 0.002890173410404624\}\}
    \end{Verbatim}

    For example, we can see that this network has 346 nodes (\emph{n}) and
938 edges (\emph{m}). The streets in this network are 11\% more
circuitous (\emph{circuity\_avg}) than straight-line would be. The
average street segment length is 121 meters
(\emph{street\_length\_avg}). We can inspect more stats, primarily
topological in nature, with the \texttt{extended\_stats} function. As
the results of many of these indicators are verbose (i.e., calculated at
the node-level), we print only the indicators' names here:

    \begin{tcolorbox}[breakable, size=fbox, boxrule=1pt, pad at break*=1mm,colback=cellbackground, colframe=cellborder]
\prompt{In}{incolor}{6}{\boxspacing}
\begin{Verbatim}[commandchars=\\\{\}]
\PY{c+c1}{\PYZsh{} calculate and print extended network stats}
\PY{n}{more\PYZus{}stats} \PY{o}{=} \PY{n}{ox}\PY{o}{.}\PY{n}{extended\PYZus{}stats}\PY{p}{(}\PY{n}{G}\PY{p}{,} \PY{n}{ecc}\PY{o}{=}\PY{k+kc}{True}\PY{p}{,} \PY{n}{bc}\PY{o}{=}\PY{k+kc}{True}\PY{p}{,} \PY{n}{cc}\PY{o}{=}\PY{k+kc}{True}\PY{p}{)}
\PY{k}{for} \PY{n}{key} \PY{o+ow}{in} \PY{n+nb}{sorted}\PY{p}{(}\PY{n}{more\PYZus{}stats}\PY{o}{.}\PY{n}{keys}\PY{p}{(}\PY{p}{)}\PY{p}{)}\PY{p}{:}
    \PY{n+nb}{print}\PY{p}{(}\PY{n}{key}\PY{p}{)}
\end{Verbatim}
\end{tcolorbox}

    \begin{Verbatim}[commandchars=\\\{\}]
avg\_neighbor\_degree
avg\_neighbor\_degree\_avg
avg\_weighted\_neighbor\_degree
avg\_weighted\_neighbor\_degree\_avg
betweenness\_centrality
betweenness\_centrality\_avg
center
closeness\_centrality
closeness\_centrality\_avg
clustering\_coefficient
clustering\_coefficient\_avg
clustering\_coefficient\_weighted
clustering\_coefficient\_weighted\_avg
degree\_centrality
degree\_centrality\_avg
diameter
eccentricity
pagerank
pagerank\_max
pagerank\_max\_node
pagerank\_min
pagerank\_min\_node
periphery
radius
    \end{Verbatim}

    The average neighborhood degree indicators refer to the mean degree of
nodes in the neighborhood of each node. The centrality indicators
(betweenness, closeness, degree, and PageRank) identify how ``central''
or important each node is to the network in terms of its topological
structure. The clustering coefficient indicators represent the extent to
which a node's neighborhood forms a complete graph. The extended stats
also include the network's eccentricity (the maximum distance from each
node to all other nodes), diameter (maximum eccentricity in the
network), radius (minimum eccentricity in the network), center (set of
all nodes whose eccentricity equals the radius), and periphery (set of
all nodes whose eccentricity equals the diameter). Additional
information about the various indicators is available online in OSMnx's
\href{https://osmnx.readthedocs.io/en/stable/osmnx.html\#module-osmnx.stats}{documentation}.

Now that we have modeled the street network and computed various
indictors of its geometry and topology, we can finally save our graph to
disk as an ESRI shapefile or a GraphML file (an open-source format for
graph serialization), allowing easy re-use in other GIS or network
analysis software:

    \begin{tcolorbox}[breakable, size=fbox, boxrule=1pt, pad at break*=1mm,colback=cellbackground, colframe=cellborder]
\prompt{In}{incolor}{7}{\boxspacing}
\begin{Verbatim}[commandchars=\\\{\}]
\PY{c+c1}{\PYZsh{} save the network model to disk as a shapefile and graphml}
\PY{n}{ox}\PY{o}{.}\PY{n}{save\PYZus{}graph\PYZus{}shapefile}\PY{p}{(}\PY{n}{G}\PY{p}{,} \PY{n}{filename}\PY{o}{=}\PY{l+s+s1}{\PYZsq{}}\PY{l+s+s1}{mynetwork\PYZus{}shapefile}\PY{l+s+s1}{\PYZsq{}}\PY{p}{)}
\PY{n}{ox}\PY{o}{.}\PY{n}{save\PYZus{}graphml}\PY{p}{(}\PY{n}{G}\PY{p}{,} \PY{n}{filename}\PY{o}{=}\PY{l+s+s1}{\PYZsq{}}\PY{l+s+s1}{mynetwork.graphml}\PY{l+s+s1}{\PYZsq{}}\PY{p}{)}
\end{Verbatim}
\end{tcolorbox}

    \hypertarget{visualizing-street-centrality}{%
\subsection{Visualizing Street
Centrality}\label{visualizing-street-centrality}}

OSMnx is built on top of NetworkX, a powerful network analysis package
developed at Los Alamos National Laboratory (Hagberg et al.~2008). We
can use it to calculate and visualize the closeness centrality of
different streets in the network. Closeness centrality measures how
central a node or edge is in a network and is defined as the reciprocal
of the sum of the distance-weighted shortest paths between the node/edge
and every other node/edge in the network.

First, we convert our graph to its line graph (sometimes called the
\emph{dual graph}; see Porta et al.~{[}2006{]}) which inverts its
topological definitions such that streets become nodes and intersections
become edges. Then we calculate the closeness centrality of each node
(i.e., street in the line graph):

    \begin{tcolorbox}[breakable, size=fbox, boxrule=1pt, pad at break*=1mm,colback=cellbackground, colframe=cellborder]
\prompt{In}{incolor}{8}{\boxspacing}
\begin{Verbatim}[commandchars=\\\{\}]
\PY{c+c1}{\PYZsh{} calculate node closeness centrality of the line graph}
\PY{n}{edge\PYZus{}centrality} \PY{o}{=} \PY{n}{nx}\PY{o}{.}\PY{n}{closeness\PYZus{}centrality}\PY{p}{(}\PY{n}{nx}\PY{o}{.}\PY{n}{line\PYZus{}graph}\PY{p}{(}\PY{n}{G}\PY{p}{)}\PY{p}{)}
\end{Verbatim}
\end{tcolorbox}

    Now that we have calculated the centrality of each street in the
network, we visualize it with matplotlib via OSMnx's
\texttt{plot\_graph} function, using the inferno color map to represent
the most-central streets in bright yellow and the least-central streets
in dark purple:

    \begin{tcolorbox}[breakable, size=fbox, boxrule=1pt, pad at break*=1mm,colback=cellbackground, colframe=cellborder]
\prompt{In}{incolor}{9}{\boxspacing}
\begin{Verbatim}[commandchars=\\\{\}]
\PY{c+c1}{\PYZsh{} make a list of graph edge centrality values}
\PY{n}{ev} \PY{o}{=} \PY{p}{[}\PY{n}{edge\PYZus{}centrality}\PY{p}{[}\PY{n}{edge} \PY{o}{+} \PY{p}{(}\PY{l+m+mi}{0}\PY{p}{,}\PY{p}{)}\PY{p}{]} \PY{k}{for} \PY{n}{edge} \PY{o+ow}{in} \PY{n}{G}\PY{o}{.}\PY{n}{edges}\PY{p}{(}\PY{p}{)}\PY{p}{]}

\PY{c+c1}{\PYZsh{} create a color scale converted to list of colors for graph edges}
\PY{n}{norm} \PY{o}{=} \PY{n}{colors}\PY{o}{.}\PY{n}{Normalize}\PY{p}{(}\PY{n}{vmin}\PY{o}{=}\PY{n+nb}{min}\PY{p}{(}\PY{n}{ev}\PY{p}{)}\PY{o}{*}\PY{l+m+mf}{0.8}\PY{p}{,} \PY{n}{vmax}\PY{o}{=}\PY{n+nb}{max}\PY{p}{(}\PY{n}{ev}\PY{p}{)}\PY{p}{)}
\PY{n}{cmap} \PY{o}{=} \PY{n}{cm}\PY{o}{.}\PY{n}{ScalarMappable}\PY{p}{(}\PY{n}{norm}\PY{o}{=}\PY{n}{norm}\PY{p}{,} \PY{n}{cmap}\PY{o}{=}\PY{n}{cm}\PY{o}{.}\PY{n}{inferno}\PY{p}{)}
\PY{n}{ec} \PY{o}{=} \PY{p}{[}\PY{n}{cmap}\PY{o}{.}\PY{n}{to\PYZus{}rgba}\PY{p}{(}\PY{n}{cl}\PY{p}{)} \PY{k}{for} \PY{n}{cl} \PY{o+ow}{in} \PY{n}{ev}\PY{p}{]}

\PY{c+c1}{\PYZsh{} color the edges in the original graph by closeness centrality in line graph}
\PY{n}{fig}\PY{p}{,} \PY{n}{ax} \PY{o}{=} \PY{n}{ox}\PY{o}{.}\PY{n}{plot\PYZus{}graph}\PY{p}{(}\PY{n}{G}\PY{p}{,} \PY{n}{bgcolor}\PY{o}{=}\PY{l+s+s1}{\PYZsq{}}\PY{l+s+s1}{black}\PY{l+s+s1}{\PYZsq{}}\PY{p}{,} \PY{n}{axis\PYZus{}off}\PY{o}{=}\PY{k+kc}{True}\PY{p}{,} \PY{n}{node\PYZus{}size}\PY{o}{=}\PY{l+m+mi}{0}\PY{p}{,}
                        \PY{n}{edge\PYZus{}color}\PY{o}{=}\PY{n}{ec}\PY{p}{,} \PY{n}{edge\PYZus{}linewidth}\PY{o}{=}\PY{l+m+mi}{2}\PY{p}{,} \PY{n}{edge\PYZus{}alpha}\PY{o}{=}\PY{l+m+mi}{1}\PY{p}{)}
\end{Verbatim}
\end{tcolorbox}

    \begin{center}
    \adjustimage{max size={0.5\linewidth}{0.5\paperheight}}{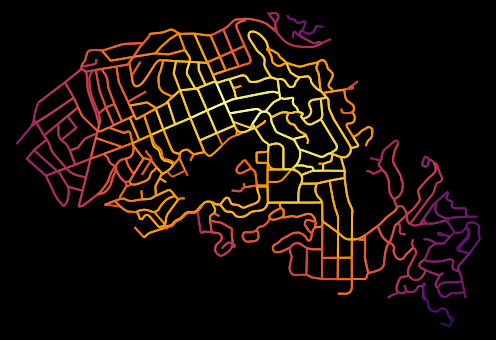}
    \end{center}
    { \hspace*{\fill} \\}
    
    \hypertarget{network-routing}{%
\subsection{Network Routing}\label{network-routing}}

OSMnx allows researchers and practitioners to calculate routes and
simulate trips along the network using various shortest-path algorithms,
such as Dijkstra's (1959). We demonstrate this here. First we use OSMnx
to find the network nodes nearest to two latitude-longitude points:

    \begin{tcolorbox}[breakable, size=fbox, boxrule=1pt, pad at break*=1mm,colback=cellbackground, colframe=cellborder]
\prompt{In}{incolor}{10}{\boxspacing}
\begin{Verbatim}[commandchars=\\\{\}]
\PY{c+c1}{\PYZsh{} find the network nodes nearest to two points}
\PY{n}{orig\PYZus{}node} \PY{o}{=} \PY{n}{ox}\PY{o}{.}\PY{n}{get\PYZus{}nearest\PYZus{}node}\PY{p}{(}\PY{n}{G}\PY{p}{,} \PY{p}{(}\PY{l+m+mf}{37.825956}\PY{p}{,} \PY{o}{\PYZhy{}}\PY{l+m+mf}{122.242278}\PY{p}{)}\PY{p}{)}
\PY{n}{dest\PYZus{}node} \PY{o}{=} \PY{n}{ox}\PY{o}{.}\PY{n}{get\PYZus{}nearest\PYZus{}node}\PY{p}{(}\PY{n}{G}\PY{p}{,} \PY{p}{(}\PY{l+m+mf}{37.817180}\PY{p}{,} \PY{o}{\PYZhy{}}\PY{l+m+mf}{122.218078}\PY{p}{)}\PY{p}{)}
\end{Verbatim}
\end{tcolorbox}

    Next we compute the shortest path between these origin and destination
nodes using Dijkstra's algorithm weighted by length (i.e., geometric
distance along the street network). Then we use OSMnx to plot this route
along the network:

    \begin{tcolorbox}[breakable, size=fbox, boxrule=1pt, pad at break*=1mm,colback=cellbackground, colframe=cellborder]
\prompt{In}{incolor}{11}{\boxspacing}
\begin{Verbatim}[commandchars=\\\{\}]
\PY{c+c1}{\PYZsh{} calculate the shortest path between these nodes then plot it}
\PY{n}{route} \PY{o}{=} \PY{n}{nx}\PY{o}{.}\PY{n}{shortest\PYZus{}path}\PY{p}{(}\PY{n}{G}\PY{p}{,} \PY{n}{orig\PYZus{}node}\PY{p}{,} \PY{n}{dest\PYZus{}node}\PY{p}{,} \PY{n}{weight}\PY{o}{=}\PY{l+s+s1}{\PYZsq{}}\PY{l+s+s1}{length}\PY{l+s+s1}{\PYZsq{}}\PY{p}{,} \PY{n}{method}\PY{o}{=}\PY{l+s+s1}{\PYZsq{}}\PY{l+s+s1}{dijkstra}\PY{l+s+s1}{\PYZsq{}}\PY{p}{)}
\PY{n}{fig}\PY{p}{,} \PY{n}{ax} \PY{o}{=} \PY{n}{ox}\PY{o}{.}\PY{n}{plot\PYZus{}graph\PYZus{}route}\PY{p}{(}\PY{n}{G}\PY{p}{,} \PY{n}{route}\PY{p}{,} \PY{n}{node\PYZus{}size}\PY{o}{=}\PY{l+m+mi}{0}\PY{p}{)}
\end{Verbatim}
\end{tcolorbox}

    \begin{center}
    \adjustimage{max size={0.5\linewidth}{0.5\paperheight}}{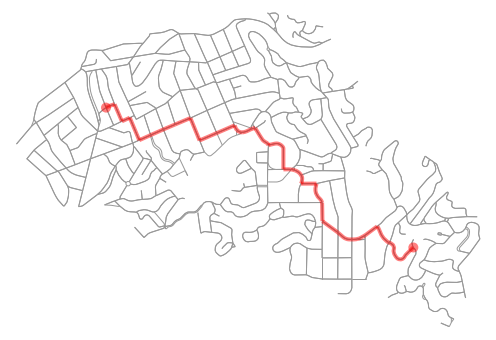}
    \end{center}
    { \hspace*{\fill} \\}
    
    Finally, we can calculate some statistics of our route, including its
total length, in meters:

    \begin{tcolorbox}[breakable, size=fbox, boxrule=1pt, pad at break*=1mm,colback=cellbackground, colframe=cellborder]
\prompt{In}{incolor}{12}{\boxspacing}
\begin{Verbatim}[commandchars=\\\{\}]
\PY{c+c1}{\PYZsh{} what is the network distance of this route?}
\PY{n}{net\PYZus{}dist} \PY{o}{=} \PY{n}{nx}\PY{o}{.}\PY{n}{shortest\PYZus{}path\PYZus{}length}\PY{p}{(}\PY{n}{G}\PY{p}{,} \PY{n}{orig\PYZus{}node}\PY{p}{,} \PY{n}{dest\PYZus{}node}\PY{p}{,} \PY{n}{weight}\PY{o}{=}\PY{l+s+s1}{\PYZsq{}}\PY{l+s+s1}{length}\PY{l+s+s1}{\PYZsq{}}\PY{p}{,} \PY{n}{method}\PY{o}{=}\PY{l+s+s1}{\PYZsq{}}\PY{l+s+s1}{dijkstra}\PY{l+s+s1}{\PYZsq{}}\PY{p}{)}
\PY{n}{net\PYZus{}dist}
\end{Verbatim}
\end{tcolorbox}

            \begin{tcolorbox}[breakable, size=fbox, boxrule=.5pt, pad at break*=1mm, opacityfill=0]
\prompt{Out}{outcolor}{12}{\boxspacing}
\begin{Verbatim}[commandchars=\\\{\}]
3284.0989999999997
\end{Verbatim}
\end{tcolorbox}
        
    Thus, this trip would travel approximately 3.3 kilometers along the
network. We can also calculate the straight-line distance between these
two network nodes as-the-crow-flies, using OSMnx's vectorized
great-circle calculator:

    \begin{tcolorbox}[breakable, size=fbox, boxrule=1pt, pad at break*=1mm,colback=cellbackground, colframe=cellborder]
\prompt{In}{incolor}{13}{\boxspacing}
\begin{Verbatim}[commandchars=\\\{\}]
\PY{c+c1}{\PYZsh{} what is the straight\PYZhy{}line distance from origin to destination?}
\PY{n}{sl\PYZus{}dist} \PY{o}{=} \PY{n}{ox}\PY{o}{.}\PY{n}{great\PYZus{}circle\PYZus{}vec}\PY{p}{(}\PY{n}{G}\PY{o}{.}\PY{n}{node}\PY{p}{[}\PY{n}{orig\PYZus{}node}\PY{p}{]}\PY{p}{[}\PY{l+s+s1}{\PYZsq{}}\PY{l+s+s1}{y}\PY{l+s+s1}{\PYZsq{}}\PY{p}{]}\PY{p}{,} \PY{n}{G}\PY{o}{.}\PY{n}{node}\PY{p}{[}\PY{n}{orig\PYZus{}node}\PY{p}{]}\PY{p}{[}\PY{l+s+s1}{\PYZsq{}}\PY{l+s+s1}{x}\PY{l+s+s1}{\PYZsq{}}\PY{p}{]}\PY{p}{,}
                              \PY{n}{G}\PY{o}{.}\PY{n}{node}\PY{p}{[}\PY{n}{dest\PYZus{}node}\PY{p}{]}\PY{p}{[}\PY{l+s+s1}{\PYZsq{}}\PY{l+s+s1}{y}\PY{l+s+s1}{\PYZsq{}}\PY{p}{]}\PY{p}{,} \PY{n}{G}\PY{o}{.}\PY{n}{node}\PY{p}{[}\PY{n}{dest\PYZus{}node}\PY{p}{]}\PY{p}{[}\PY{l+s+s1}{\PYZsq{}}\PY{l+s+s1}{x}\PY{l+s+s1}{\PYZsq{}}\PY{p}{]}\PY{p}{)}
\PY{n}{sl\PYZus{}dist}
\end{Verbatim}
\end{tcolorbox}

            \begin{tcolorbox}[breakable, size=fbox, boxrule=.5pt, pad at break*=1mm, opacityfill=0]
\prompt{Out}{outcolor}{13}{\boxspacing}
\begin{Verbatim}[commandchars=\\\{\}]
2340.8766018171827
\end{Verbatim}
\end{tcolorbox}
        
    Comparing these two distance values, we can compute an indicator of trip
circuity: that is, how much greater the network-constrained distance is
between two nodes compared to the straight-line distance between them.
In this case, we can see that the network distance is approximately 40\%
longer than the straight-line distance:

    \begin{tcolorbox}[breakable, size=fbox, boxrule=1pt, pad at break*=1mm,colback=cellbackground, colframe=cellborder]
\prompt{In}{incolor}{14}{\boxspacing}
\begin{Verbatim}[commandchars=\\\{\}]
\PY{c+c1}{\PYZsh{} how much longer is the network distance than the straight\PYZhy{}line?}
\PY{n}{net\PYZus{}dist} \PY{o}{/} \PY{n}{sl\PYZus{}dist}
\end{Verbatim}
\end{tcolorbox}

            \begin{tcolorbox}[breakable, size=fbox, boxrule=.5pt, pad at break*=1mm, opacityfill=0]
\prompt{Out}{outcolor}{14}{\boxspacing}
\begin{Verbatim}[commandchars=\\\{\}]
1.4029355487814306
\end{Verbatim}
\end{tcolorbox}
        
    \hypertarget{downloadingmodeling-networks-in-other-ways}{%
\subsection{Downloading/Modeling Networks in Other
Ways}\label{downloadingmodeling-networks-in-other-ways}}

So far, we have modeled and analyzed the street network of Piedmont,
California. However, we are not constrained to study sites in the United
States. OpenStreetMap is a global mapping project and OSMnx can model
networks anywhere in the world, such as Modena, Italy:

    \begin{tcolorbox}[breakable, size=fbox, boxrule=1pt, pad at break*=1mm,colback=cellbackground, colframe=cellborder]
\prompt{In}{incolor}{15}{\boxspacing}
\begin{Verbatim}[commandchars=\\\{\}]
\PY{c+c1}{\PYZsh{} create a graph of Modena\PYZsq{}s drivable street network then plot it}
\PY{n}{G} \PY{o}{=} \PY{n}{ox}\PY{o}{.}\PY{n}{graph\PYZus{}from\PYZus{}place}\PY{p}{(}\PY{l+s+s1}{\PYZsq{}}\PY{l+s+s1}{Modena, Italy}\PY{l+s+s1}{\PYZsq{}}\PY{p}{,} \PY{n}{retain\PYZus{}all}\PY{o}{=}\PY{k+kc}{True}\PY{p}{)}
\PY{n}{fig}\PY{p}{,} \PY{n}{ax} \PY{o}{=} \PY{n}{ox}\PY{o}{.}\PY{n}{plot\PYZus{}graph}\PY{p}{(}\PY{n}{G}\PY{p}{,} \PY{n}{fig\PYZus{}height}\PY{o}{=}\PY{l+m+mi}{8}\PY{p}{,} \PY{n}{node\PYZus{}size}\PY{o}{=}\PY{l+m+mi}{0}\PY{p}{,} \PY{n}{edge\PYZus{}linewidth}\PY{o}{=}\PY{l+m+mf}{0.5}\PY{p}{)}
\end{Verbatim}
\end{tcolorbox}

    \begin{center}
    \adjustimage{max size={0.5\linewidth}{0.5\paperheight}}{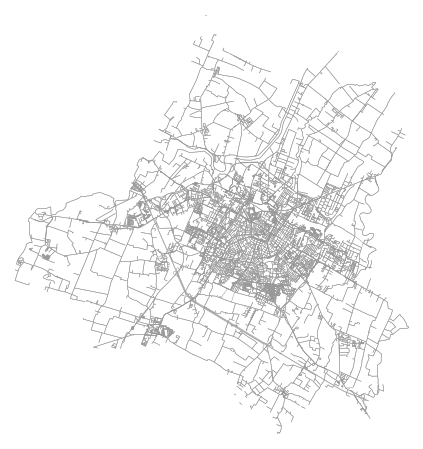}
    \end{center}
    { \hspace*{\fill} \\}
    
    We have seen how to download street network data and turn it into a
graph-based model using OSMnx's \texttt{graph\_from\_place} function.
This function geocodes the place name using OpenStreetMap's Nominatim
web service, identifies its bounding polygon, then downloads all the
network data within this polygon from OpenStreetMap's Overpass API. This
workflow easily handles well-defined place names. However, OSMnx offers
additional functionality to download and model networks for other study
sites as well.

For example, if OpenStreetMap does not have a bounding polygon for a
specific study site, we can acquire its street network anyway by passing
a polygon directly into the \texttt{graph\_from\_polygon} function. Or
we can pass in latitude-longitude coordinates and a distance into the
\texttt{graph\_from\_point} function as demonstrated here, where we
visualize the network within a bounding box around the University of
California, Berkeley's Wurster Hall:

    \begin{tcolorbox}[breakable, size=fbox, boxrule=1pt, pad at break*=1mm,colback=cellbackground, colframe=cellborder]
\prompt{In}{incolor}{16}{\boxspacing}
\begin{Verbatim}[commandchars=\\\{\}]
\PY{c+c1}{\PYZsh{} create a graph around UC Berkeley then plot it}
\PY{n}{wurster\PYZus{}hall} \PY{o}{=} \PY{p}{(}\PY{l+m+mf}{37.870605}\PY{p}{,} \PY{o}{\PYZhy{}}\PY{l+m+mf}{122.254830}\PY{p}{)}
\PY{n}{one\PYZus{}mile} \PY{o}{=} \PY{l+m+mi}{1609} \PY{c+c1}{\PYZsh{}one mile in meters}
\PY{n}{G} \PY{o}{=} \PY{n}{ox}\PY{o}{.}\PY{n}{graph\PYZus{}from\PYZus{}point}\PY{p}{(}\PY{n}{wurster\PYZus{}hall}\PY{p}{,} \PY{n}{distance}\PY{o}{=}\PY{n}{one\PYZus{}mile}\PY{p}{,} \PY{n}{network\PYZus{}type}\PY{o}{=}\PY{l+s+s1}{\PYZsq{}}\PY{l+s+s1}{drive}\PY{l+s+s1}{\PYZsq{}}\PY{p}{)}
\PY{n}{fig}\PY{p}{,} \PY{n}{ax} \PY{o}{=} \PY{n}{ox}\PY{o}{.}\PY{n}{plot\PYZus{}graph}\PY{p}{(}\PY{n}{G}\PY{p}{,} \PY{n}{node\PYZus{}size}\PY{o}{=}\PY{l+m+mi}{0}\PY{p}{)}
\end{Verbatim}
\end{tcolorbox}

    \begin{center}
    \adjustimage{max size={0.5\linewidth}{0.5\paperheight}}{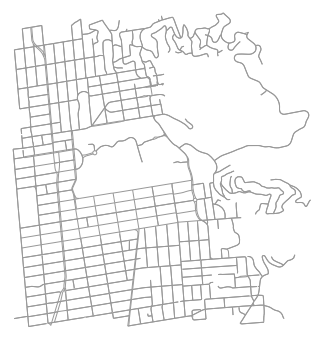}
    \end{center}
    { \hspace*{\fill} \\}
    
    OSMnx also accepts place queries as unambiguous Python dictionaries to
help the geocoder find a specific matching study site when several names
might approximately overlap. In this example, we download the street
network of San Francisco, California by defining the query with such a
dictionary:

    \begin{tcolorbox}[breakable, size=fbox, boxrule=1pt, pad at break*=1mm,colback=cellbackground, colframe=cellborder]
\prompt{In}{incolor}{17}{\boxspacing}
\begin{Verbatim}[commandchars=\\\{\}]
\PY{c+c1}{\PYZsh{} create a graph of San Francisco\PYZsq{}s drivable street network then plot it}
\PY{n}{place} \PY{o}{=} \PY{p}{\PYZob{}}\PY{l+s+s1}{\PYZsq{}}\PY{l+s+s1}{city}\PY{l+s+s1}{\PYZsq{}}   \PY{p}{:} \PY{l+s+s1}{\PYZsq{}}\PY{l+s+s1}{San Francisco}\PY{l+s+s1}{\PYZsq{}}\PY{p}{,}
         \PY{l+s+s1}{\PYZsq{}}\PY{l+s+s1}{state}\PY{l+s+s1}{\PYZsq{}}  \PY{p}{:} \PY{l+s+s1}{\PYZsq{}}\PY{l+s+s1}{California}\PY{l+s+s1}{\PYZsq{}}\PY{p}{,}
         \PY{l+s+s1}{\PYZsq{}}\PY{l+s+s1}{country}\PY{l+s+s1}{\PYZsq{}}\PY{p}{:} \PY{l+s+s1}{\PYZsq{}}\PY{l+s+s1}{USA}\PY{l+s+s1}{\PYZsq{}}\PY{p}{\PYZcb{}}
\PY{n}{G} \PY{o}{=} \PY{n}{ox}\PY{o}{.}\PY{n}{graph\PYZus{}from\PYZus{}place}\PY{p}{(}\PY{n}{place}\PY{p}{,} \PY{n}{network\PYZus{}type}\PY{o}{=}\PY{l+s+s1}{\PYZsq{}}\PY{l+s+s1}{drive}\PY{l+s+s1}{\PYZsq{}}\PY{p}{)}
\end{Verbatim}
\end{tcolorbox}

    \hypertarget{downloading-other-infrastructure-types}{%
\subsection{Downloading Other Infrastructure
Types}\label{downloading-other-infrastructure-types}}

All of the preceding examples have focused on urban and suburban street
networks. However, OSMnx can also download and model other networked
infrastructure types by passing in custom queries via the
\texttt{infrastructure} argument. Such networked infrastructure could
include power lines, the canal systems of Venice or Amsterdam, or the
New York City subway's rail infrastructure as illustrated in this
example:

    \begin{tcolorbox}[breakable, size=fbox, boxrule=1pt, pad at break*=1mm,colback=cellbackground, colframe=cellborder]
\prompt{In}{incolor}{18}{\boxspacing}
\begin{Verbatim}[commandchars=\\\{\}]
\PY{c+c1}{\PYZsh{} create a graph of NYC\PYZsq{}s subway rail infrastructure then plot it}
\PY{n}{G} \PY{o}{=} \PY{n}{ox}\PY{o}{.}\PY{n}{graph\PYZus{}from\PYZus{}place}\PY{p}{(}\PY{l+s+s1}{\PYZsq{}}\PY{l+s+s1}{New York City, New York, USA}\PY{l+s+s1}{\PYZsq{}}\PY{p}{,}
                        \PY{n}{retain\PYZus{}all}\PY{o}{=}\PY{k+kc}{False}\PY{p}{,} \PY{n}{truncate\PYZus{}by\PYZus{}edge}\PY{o}{=}\PY{k+kc}{True}\PY{p}{,} \PY{n}{simplify}\PY{o}{=}\PY{k+kc}{True}\PY{p}{,}
                        \PY{n}{network\PYZus{}type}\PY{o}{=}\PY{l+s+s1}{\PYZsq{}}\PY{l+s+s1}{none}\PY{l+s+s1}{\PYZsq{}}\PY{p}{,} \PY{n}{infrastructure}\PY{o}{=}\PY{l+s+s1}{\PYZsq{}}\PY{l+s+s1}{way[}\PY{l+s+s1}{\PYZdq{}}\PY{l+s+s1}{railway}\PY{l+s+s1}{\PYZdq{}}\PY{l+s+s1}{\PYZti{}}\PY{l+s+s1}{\PYZdq{}}\PY{l+s+s1}{subway}\PY{l+s+s1}{\PYZdq{}}\PY{l+s+s1}{]}\PY{l+s+s1}{\PYZsq{}}\PY{p}{)}

\PY{n}{fig}\PY{p}{,} \PY{n}{ax} \PY{o}{=} \PY{n}{ox}\PY{o}{.}\PY{n}{plot\PYZus{}graph}\PY{p}{(}\PY{n}{G}\PY{p}{,} \PY{n}{node\PYZus{}size}\PY{o}{=}\PY{l+m+mi}{0}\PY{p}{)}
\end{Verbatim}
\end{tcolorbox}

    \begin{center}
    \adjustimage{max size={0.35\linewidth}{0.35\paperheight}}{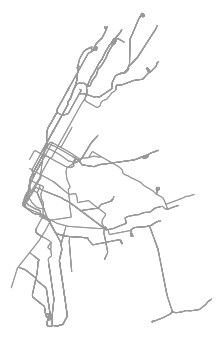}
    \end{center}
    { \hspace*{\fill} \\}
    
    Note that the preceding code snippet modeled subway rail
\emph{infrastructure} which thus includes crossovers, sidings, spurs,
yards, and the like. For a station-based train network model, the
analyst would be best-served downloading and modeling a station
adjacency matrix.

Beyond networked infrastructure, OSMnx can also work with OpenStreetMap
building footprint and points of interest data. For example, we can
download and visualize the building footprints near New York's Empire
State Building:

    \begin{tcolorbox}[breakable, size=fbox, boxrule=1pt, pad at break*=1mm,colback=cellbackground, colframe=cellborder]
\prompt{In}{incolor}{19}{\boxspacing}
\begin{Verbatim}[commandchars=\\\{\}]
\PY{c+c1}{\PYZsh{} download and visualize the building footprints around the empire state bldg}
\PY{n}{point} \PY{o}{=} \PY{p}{(}\PY{l+m+mf}{40.748482}\PY{p}{,} \PY{o}{\PYZhy{}}\PY{l+m+mf}{73.985402}\PY{p}{)} \PY{c+c1}{\PYZsh{}empire state bldg coordinates}
\PY{n}{dist} \PY{o}{=} \PY{l+m+mi}{812} \PY{c+c1}{\PYZsh{}meters}
\PY{n}{gdf} \PY{o}{=} \PY{n}{ox}\PY{o}{.}\PY{n}{footprints\PYZus{}from\PYZus{}point}\PY{p}{(}\PY{n}{point}\PY{o}{=}\PY{n}{point}\PY{p}{,} \PY{n}{distance}\PY{o}{=}\PY{n}{dist}\PY{p}{)}
\PY{n}{gdf\PYZus{}proj} \PY{o}{=} \PY{n}{ox}\PY{o}{.}\PY{n}{project\PYZus{}gdf}\PY{p}{(}\PY{n}{gdf}\PY{p}{)}
\PY{n}{bbox\PYZus{}proj} \PY{o}{=} \PY{n}{ox}\PY{o}{.}\PY{n}{bbox\PYZus{}from\PYZus{}point}\PY{p}{(}\PY{n}{point}\PY{o}{=}\PY{n}{point}\PY{p}{,} \PY{n}{distance}\PY{o}{=}\PY{n}{dist}\PY{p}{,} \PY{n}{project\PYZus{}utm}\PY{o}{=}\PY{k+kc}{True}\PY{p}{)}
\PY{n}{fig}\PY{p}{,} \PY{n}{ax} \PY{o}{=} \PY{n}{ox}\PY{o}{.}\PY{n}{plot\PYZus{}footprints}\PY{p}{(}\PY{n}{gdf\PYZus{}proj}\PY{p}{,} \PY{n}{bbox}\PY{o}{=}\PY{n}{bbox\PYZus{}proj}\PY{p}{,} \PY{n}{bgcolor}\PY{o}{=}\PY{l+s+s1}{\PYZsq{}}\PY{l+s+s1}{\PYZsh{}333333}\PY{l+s+s1}{\PYZsq{}}\PY{p}{,} \PY{n}{color}\PY{o}{=}\PY{l+s+s1}{\PYZsq{}}\PY{l+s+s1}{w}\PY{l+s+s1}{\PYZsq{}}\PY{p}{,} \PY{n}{figsize}\PY{o}{=}\PY{p}{(}\PY{l+m+mi}{6}\PY{p}{,}\PY{l+m+mi}{6}\PY{p}{)}\PY{p}{)}
\end{Verbatim}
\end{tcolorbox}

    \begin{center}
    \adjustimage{max size={0.4\linewidth}{0.4\paperheight}}{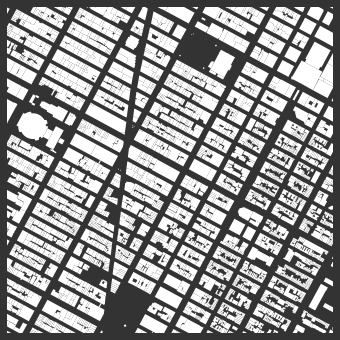}
    \end{center}
    { \hspace*{\fill} \\}
    
    Finally, we can download and inspect the amenities matching the tag
``restaurants'' near the Empire State Building and then display the five
most common cuisine types

    \begin{tcolorbox}[breakable, size=fbox, boxrule=1pt, pad at break*=1mm,colback=cellbackground, colframe=cellborder]
\prompt{In}{incolor}{20}{\boxspacing}
\begin{Verbatim}[commandchars=\\\{\}]
\PY{c+c1}{\PYZsh{} download restaurants near the empire state bldg then display them}
\PY{n}{gdf} \PY{o}{=} \PY{n}{ox}\PY{o}{.}\PY{n}{pois\PYZus{}from\PYZus{}point}\PY{p}{(}\PY{n}{point}\PY{o}{=}\PY{n}{point}\PY{p}{,} \PY{n}{distance}\PY{o}{=}\PY{n}{dist}\PY{p}{,} \PY{n}{amenities}\PY{o}{=}\PY{p}{[}\PY{l+s+s1}{\PYZsq{}}\PY{l+s+s1}{restaurant}\PY{l+s+s1}{\PYZsq{}}\PY{p}{]}\PY{p}{)}
\PY{n}{gdf}\PY{p}{[}\PY{p}{[}\PY{l+s+s1}{\PYZsq{}}\PY{l+s+s1}{name}\PY{l+s+s1}{\PYZsq{}}\PY{p}{,} \PY{l+s+s1}{\PYZsq{}}\PY{l+s+s1}{cuisine}\PY{l+s+s1}{\PYZsq{}}\PY{p}{]}\PY{p}{]}\PY{o}{.}\PY{n}{dropna}\PY{p}{(}\PY{p}{)}\PY{o}{.}\PY{n}{head}\PY{p}{(}\PY{p}{)}
\end{Verbatim}
\end{tcolorbox}

            \begin{tcolorbox}[breakable, size=fbox, boxrule=.5pt, pad at break*=1mm, opacityfill=0]
\prompt{Out}{outcolor}{20}{\boxspacing}
\begin{Verbatim}[commandchars=\\\{\}]
                                name         cuisine
357620442  Dolcino Trattoria Toscana         italian
419359995               Little Alley         chinese
419367625               Ramen Takumi  japanese;ramen
561042187                 Les Halles          french
663104998            Tick Tock Diner           diner
\end{Verbatim}
\end{tcolorbox}
        
    \begin{tcolorbox}[breakable, size=fbox, boxrule=1pt, pad at break*=1mm,colback=cellbackground, colframe=cellborder]
\prompt{In}{incolor}{21}{\boxspacing}
\begin{Verbatim}[commandchars=\\\{\}]
\PY{c+c1}{\PYZsh{} show the five most common cuisine types among these restaurants}
\PY{n}{gdf}\PY{p}{[}\PY{l+s+s1}{\PYZsq{}}\PY{l+s+s1}{cuisine}\PY{l+s+s1}{\PYZsq{}}\PY{p}{]}\PY{o}{.}\PY{n}{value\PYZus{}counts}\PY{p}{(}\PY{p}{)}\PY{o}{.}\PY{n}{head}\PY{p}{(}\PY{p}{)}
\end{Verbatim}
\end{tcolorbox}

            \begin{tcolorbox}[breakable, size=fbox, boxrule=.5pt, pad at break*=1mm, opacityfill=0]
\prompt{Out}{outcolor}{21}{\boxspacing}
\begin{Verbatim}[commandchars=\\\{\}]
indian      22
korean      15
italian     14
japanese    13
pizza        9
Name: cuisine, dtype: int64
\end{Verbatim}
\end{tcolorbox}
        
    \hypertarget{conclusion}{%
\section{Conclusion}\label{conclusion}}

This article argued that computational notebooks underpin an important
emerging pillar in urban analytics and planning research, pedagogy, and
practice. To demonstrate this, it presented the official repository of
computational notebooks that the OSMnx project uses for tutorials,
demos, and guides. It illustrated the use of these notebooks by
highlighting specific examples from them, inline and interactively
within this article, as an introduction to this modeling and analysis
software. OSMnx itself is a Python package for downloading, modeling,
analyzing, and visualizing data from OpenStreetMap. It lets users
analyze networked infrastructure like street networks as well as
building footprints, points of interest, elevation data, and more. This
article demonstrated how computational notebooks can provide a
tutorial-style introduction to scientific software such as this.

The OSMnx project uses computational notebooks because they offer
several advantages. First, they empower scientific reproducibility,
replication, sharing, and remixing. Second, they allow researchers to
intermingle data analyses with visualizations and narratives to ask and
answer research questions. Third, they offer ``follow-along'' guides for
introducing software and methods to new users, such as in this
repository for OSMnx or even in the university classroom. Finally, they
help researchers reach a wider community of interest by making their
methodologies and analyses more legible to a broad audience potentially
interested in adapting and remixing their work. For these reasons and
more, we expect to see growing adoption of computational notebooks in
the urban planning discipline and related analytics fields.

    \hypertarget{appendix}{%
\section*{Appendix}\label{appendix}}

The interested reader may consult the following web sites for more
information and resources as discussed in the article:

\begin{itemize}
\tightlist
\item
  OSMnx examples repository: https://github.com/gboeing/osmnx-examples
\item
  OSMnx documentation: https://osmnx.readthedocs.io/
\item
  Docker Desktop is available at:
  https://www.docker.com/products/docker-desktop
\item
  The OSMnx Docker image is available at:
  https://hub.docker.com/r/gboeing/osmnx
\end{itemize}

    \hypertarget{references}{%
\section*{References}\label{references}}

Batty, M. (2019). Urban Analytics Defined. \emph{Environment and
Planning B: Urban Analytics and City Science}, 46(3), 403--405.
https://doi.org/10.1177/2399808319839494

Boeing, G. (2017). OSMnx: New Methods for Acquiring, Constructing,
Analyzing, and Visualizing Complex Street Networks. \emph{Computers,
Environment and Urban Systems}, 65, 126--139.
https://doi.org/10.1016/j.compenvurbsys.2017.05.004

Dijkstra, E. W. (1959). A Note on Two Problems in Connexion with Graphs.
\emph{Numerische Mathematik}, 1(1), 269--271.
https://doi.org/10.1007/BF01386390

Hagberg, A. A., Schult, D. A., \& Swart, P. J. (2008). Exploring Network
Structure, Dynamics, and Function using NetworkX. In G. Varoquaux, T.
Vaught, \& J. Millman (Eds.), \emph{Proceedings of the 7th Python in
Science Conference} (pp.~11--16). Pasadena, California: SciPy.

Harris, R., O'Sullivan, D., Gahegan, M., Charlton, M., Comber, L.,
Longley, P., \ldots{} Evans, A. (2017). More Bark than Bytes?
Reflections on 21+ Years of Geocomputation. \emph{Environment and
Planning B: Urban Analytics and City Science}, 44(4), 598--617.
https://doi.org/10.1177/2399808317710132

Kang, W., Oshan, T., Wolf, L.J., Boeing, G., Frias-Martinez, V., Gao,
S., Poorthuis, A., Xu, W., 2019. A Roundtable Discussion: Defining Urban
Data Science. \emph{Environment and Planning B: Urban Analytics and City
Science}, 46(9), 1756--1768. https://doi.org/10.1177/2399808319882826

Kedron, P., Frazier, A.E., Trgovac, A.B., Nelson, T., Fotheringham,
A.S., 2019. Reproducibility and Replicability in Geographical Analysis.
\emph{Geographical Analysis}, published online ahead of print.
https://doi.org/10.1111/gean.12221

Kluyver, T., Ragan-Kelley, B., Pérez, F., Granger, B., Bussonnier, M.,
Frederic, J., \ldots{} Jupyter Development Team. (2016). Jupyter
Notebooks: A Publishing Format for Reproducible Computational Workflows.
In F. Loizides \& B. Schmidt (Eds.), \emph{Positioning and Power in
Academic Publishing: Players, Agents and Agendas} (pp.~87--90).
Amsterdam, Netherlands: IOS Press.
https://doi.org/10.3233/978-1-61499-649-1-87

Knuth, D. E. (1992). \emph{Literate Programming}. Stanford, CA: Center
for the Study of Language and Information.

Kontokosta, C. E. (2018). Urban Informatics in the Science and Practice
of Planning. \emph{Journal of Planning Education and Research},
published online ahead of print.
https://doi.org/10.1177/0739456X18793716

Padgham, M., Boeing, G., Cooley, D., Tierney, N., Sumner, M., Phan,
T.G., Beare, R., 2019. An Introduction to Software Tools, Data, and
Services for Geospatial Analysis of Stroke Services. \emph{Frontiers in
Neurology}, 10, 743. https://doi.org/10.3389/fneur.2019.00743

Pérez, F., \& Granger, B. E. (2007). IPython: A System for Interactive
Scientific Computing. \emph{Computing in Science \& Engineering}, 9(3),
21--29. https://doi.org/10.1109/MCSE.2007.53

Poorthuis, A., \& Zook, M. (2019). Being Smarter about Space: Drawing
Lessons from Spatial Science. \emph{Annals of the American Association
of Geographers}, published online ahead of print.
https://doi.org/10.1080/24694452.2019.1674630

Porta, S., Crucitti, P., \& Latora, V. (2006). The Network Analysis of
Urban Streets: A Dual Approach. \emph{Physica A: Statistical Mechanics
and Its Applications}, 369(2), 853--866.
https://doi.org/10.1016/j.physa.2005.12.063

Rey, S. J. (2019). PySAL: The First 10 Years. \emph{Spatial Economic
Analysis}, 14(3), 273--282.
https://doi.org/10.1080/17421772.2019.1593495

Singleton, A., Arribas‐Bel, D., 2019. Geographic Data Science.
\emph{Geographical Analysis}, published online ahead of print.
https://doi.org/10.1111/gean.12194


\end{document}